\documentclass[journal]{IEEEtran}

\usepackage{xcolor, soul, framed, tabularx, algorithm, 
 algcompatible, amsmath} 
\usepackage{multirow}

\colorlet{shadecolor}{yellow}
\usepackage[pdftex]{graphicx}
\graphicspath{{../pdf/}{../jpeg/}}
\DeclareGraphicsExtensions{.pdf,.jpeg,.png}

\usepackage{amsmath}
\usepackage{array}
\usepackage{dsfont}
\usepackage{bbold}
\usepackage{mdwmath}
\usepackage{mdwtab}
\usepackage{eqparbox}
\usepackage{url}

\begin{document}
\title{SDDPM: Speckle Denoising Diffusion Probabilistic Models}

  \author{Soumee~Guha,~\IEEEmembership{Student Member,~IEEE,}
      and~Scott~T.~Acton,~\IEEEmembership{Fellow,~IEEE}

  \thanks{S. Guha and S. T. Acton are with the Department of Electrical and  Computer Engineering, University of Virginia, VA, 22904-4743 USA (e-mail: ccf3dv@virginia.edu; acton@virginia.edu). This work was funded in part by NIH under 1R01GM139002 and in part by the ARO under W911NF2010206.}
}

\markboth{IEEE TRANSACTION ON IMAGE PROCESSING
}{Roberg \MakeLowercase{\textit{et al.}}: SDDPM: Speckle Denoising Diffusion Probabilistic Model}

\maketitle

\begin{abstract}
Coherent imaging systems, such as medical ultrasound and synthetic aperture radar (SAR), are subject to corruption from speckle due to sub-resolution scatterers. Since speckle is multiplicative in nature, the constituent image regions become corrupted to different extents. The task of denoising such images requires algorithms specifically designed for removing signal-dependent noise. This paper proposes a novel image-denoising algorithm for removing signal-dependent multiplicative noise with diffusion models - Speckle Denoising Diffusion Probabilistic Models (SDDPM). We derive the mathematical formulations for the forward process, the reverse process, and the training objective. In the forward process, we apply multiplicative noise to a given image and prove that the forward process is Gaussian. We show that the reverse process is also Gaussian and the final training objective can be expressed as the Kullback–Leibler (KL) divergence between the forward and reverse processes. As derived in the paper, the final denoising task is a single-step process, thereby reducing the denoising time significantly. We have trained our model with natural land-use images and ultrasound images for different noise levels. Extensive experiments centered around two different applications show that SDDPM is robust and performs significantly better than the comparative models even when the images are severely corrupted. 
\end{abstract}


%
\IEEEpeerreviewmaketitle


\section{Introduction}

Image denoising has been studied by several researchers over the past few decades \cite{yu2002speckle, dabov2006image, ramani2008monte, zhang2017beyond}. Noise in images can be attributed to different methodologies involved with the acquisition, compression, and transmission processes, leading to significant distortion and loss of important information \cite{fan2019brief}. Image denoising can be defined as an inverse problem that involves retrieving a denoised version of the image while preserving the important image features and improving the signal-to-noise ratio. Different types of images are corrupted with varying kinds of noise, i.e., Gaussian noise, Poisson noise, speckle, salt and pepper noise, etc. Different types of noise sources have different characteristics and taking into account the specific noise model can lead to efficient image denoising algorithms. 

Ultrasound images and synthetic aperture radar (SAR) images are plagued by speckle. Unlike many other types of noise, speckle is signal-dependent. Thus, subsequent tasks like image segmentation and image classification become challenging. Over the last few decades, many filtering algorithms have been designed for speckle removal \cite{frost1982model, kuan1983adaptive, lee1980digital}. The filtering algorithms are typically sensitive to the shape of the filter window and the filter size, often leading to oversmoothing or artifacts in the denoised images. \cite{yu2002speckle} proposed a partial differential equation (PDE)-based approach for denoising images plagued by speckle noise. The denoised images not only preserved the edges but also enhanced them. However, all these methods are iterative and require a significant amount of computational expense to generate results. 

Alongside classical image denoising methods, a host of neural network models have also been proposed in recent years. A neural network-based image denoising model can be trained to generate the denoised version of any input image. In the past few years, denoising diffusion probabilistic models (DDPM), introduced by \cite{sohl2015deep, ho2020denoising} have been extremely successful for a wide range of applications. A diffusion model is characterized by a forward and reverse process and is parameterized as a Markov chain. The forward process involves gradually adding noise until the initial image is absolutely corrupted. In the reverse process, a neural network model is trained to denoise the corrupted image in small steps. Diffusion models can generate remarkable results in both conditional and unconditional image generation. 

To date, the denoising methods proposed using DDPM have assumed an additive model. Unlike additive noise, multiplicative noise (speckle) corrupts different parts of the image differently according to intensity. Mathematically, when an input image $I$ is corrupted by signal-dependent noise $N$, the corrupted image $I'$ can be represented as:
\begin{equation}
    \label{speckle_noise_I}
    I' = I + I \, N
\end{equation}
In (\ref{speckle_noise_I}), $N$ is zero-mean Gaussian with some variance, i.e., 
\begin{equation}
    N \sim \mathcal{N}(\mathbf{0}, \sigma^2)
\end{equation}
The variance of $N$ determines the amount of noise present in the image. In the realm of image processing, additive noise induces a translation in the pixel intensities of an image. It achieves this by introducing random values that are either added or subtracted to the original pixel values, thereby instigating a persistent bias within the observations. Conversely, multiplicative noise operates by adjusting the original pixel values through a scaling mechanism. This form of noise is notably predisposed to influencing both the variance and the interdependencies among the pixel intensities in the image. Thus, compared to additive noise, speckle noise degrades images more severely, thereby making the downstream tasks further complicated. 

In this work, we have designed a diffusion model for denoising images corrupted with speckle. To the best of our knowledge, this is the first work that attempts to implement a diffusion model that considers multiplicative noise instead of additive Gaussian noise. \cite{Perera_2023} has proposed  SAR-DDPM which is aimed at removing multiplicative noise from SAR images with the denoising diffusion probabilistic model (DDPM) \cite{ho2020denoising}. However, our work is fundamentally different from the SAR-DDPM model proposed in \cite{Perera_2023} as their model utilizes additive noise in the forward and reverse processes. The architecture used by SAR-DDPM is also different from ours. In SAR-DDPM, the diffusion model is given both the noisy and the original images and the denoising process is performed in \textit{T} timesteps. In contrast, the proposed method assumes a multiplicative noise model, and the denoising task is a single-step process. The details are mentioned in Section \ref{noise_model}. The main contributions of this work are: \\
\begin{itemize}
    \item We propose a novel model Speckle Denoising Diffusion Probabilistic Models (SDDPM) for denoising speckled images that is particularly attuned to multiplicative noise. Both the forward and reverse processes assume multiplicative signal-dependent speckle. \\
    \item We derive the training objective for generating the despeckled given any image corrupted with speckle. \\
    \item Extensive experiments with different datasets show that our model achieves superior performance with respect to the state-of-the-art image denoising algorithms.
\end{itemize}

\section{Literature Survey}

\subsection{Speckle Removal Methods}

The Lee \cite{lee1980digital}, Frost \cite{frost1982model} and Kuan \cite{kuan1983adaptive} filters are the early solutions for removing speckle. For the Lee and Kuan filters, the denoised image is a combination of the intensity of the center pixel in a window and the average intensity for the window. These filters smooth the homogeneous regions by averaging while attempting to preserve edges and point structures. The Frost filter is a linear, convolutional filter. It is adaptive in nature and is a circularly symmetric filter with exponential damping. In speckle reducing anisotropic diffusion (SRAD) \cite{yu2002speckle}, the authors propose a partial differential equation-based speckle removal method which not only preserves the existing edges but also enhances them. SRAD is the edge-sensitive extension of the contemporary adaptive speckle removal methods. \cite{gupta2004wavelet} introduces a novel speckle reduction algorithm by soft thresholding of the wavelet coefficients of the logarithm of the speckled image. The wavelet coefficients of the noiseless signal are modeled with a generalized Gaussian distribution whereas the speckle is modeled with a Gaussian distribution. \cite{coupe2009nonlocal} use a Bayesian framework to derive an adaptation of the non-local means filter which is adapted to the noise model of the ultrasound images and uses Pearson distance for comparing different image patches. The classical bilateral filter is efficient for removing Gaussian noise while preserving edges. \cite{singh2020local} propose a real-time speckle-removing filter using local statistics of the images in the bilateral filter framework. A cluster-based speckle reduction technique is proposed in \cite{eybposh2018cluster} which attempts to remove speckle noise in two steps. In order to preserve the edges in the denoised image, the pixels are first clustered for detecting the edges and thereafter, an adaptive filtering method is applied to the different pixel clusters for noise removal. 

There exist a number of speckle removing algorithms from the deep neural network community as well. An adversarial framework has been proposed in \cite{bobrow2019deeplsr} for laser speckle removal in which images having coherent illumination are transformed to speckle-free images having incoherent illumination. For despeckling SAR images, access to time series data representing different speckle realizations of the same area can be beneficial. \cite{meraoumia2023multitemporal} utilizes this idea and has proposed a self-supervised deep framework for generating a single despeckled SAR image given a time series of coregistered inputs. The self-supervised framework eliminates the need for a large volume of ground truth images. To eliminate the need for high quality well-registered noisy and clean image pairs, an unsupervised learning framework has been proposed in \cite{huang2020noise}. In this work, noisy images are first decomposed into the latent representations of the denoised content and the noise. Then, a generative framework is used to predict the final despeckled images from the latent representation of the denoised content. An edge-preserving speckle reduction framework has been proposed in \cite{ma2018speckle}. This solution implements a conditional generative adversarial network (cGAN) having a separate component for the edge loss along with the objective function for the cGAN thereby an edge-sensitive optimizing function.

\subsection{Diffusion Models}

Diffusion models were first introduced in \cite{sohl2015deep, ho2020denoising} as a latent variable model. They 
are used to convert a sample from a Gaussian distribution to an arbitrarily complex target distribution. Mathematically, diffusion models can be defined by a forward process $q(\cdot)$ and a reverse process $p_{\theta}(\cdot)$, both of which maintain the Markov property. Starting with image $x_0$, the forward process $q$ adds Gaussian noise in \textit{T} steps according to a noise schedule $\alpha_t$ where $t \in [1, T].$ The noise parameters are designed such that $x_T \sim \mathcal{N}(0, I)$. The reverse process $p$ involves denoising $x_T$ in \textit{T} iterations.

Formally, the forward process can be written as:
\begin{equation}
    \label{additivenoise}
    x_t = \sqrt{\alpha_t}x_{t-1} + \sqrt{1 - \alpha_t} \epsilon; \epsilon \sim \mathcal{N}(\epsilon; \mathbf{0}, \mathbf{I})
\end{equation}

\begin{equation}
    q(x_t| x_{t-1}) = \mathcal{N}(x_t; \sqrt{\alpha_t}x_{t-1}, (1-\alpha_t)\mathbf{I}) 
\end{equation}

\begin{equation}
    q(x_{1:T}| x_0) = \prod_{t=1}^T q(x_t|x_{t-1})
\end{equation}
\\
The reverse process $p$ can be expressed as:
\begin{equation}
    p(x_{0:T}) = p(x_T) \prod_{t=1}^T p_\theta(x_{t-1}|x_t)
\end{equation}

Each step of the denoising process is learned by a neural network parameterized by $\theta$ and can be simplified as:

\begin{equation}
    p_\theta (x_{t-1}|x_t) = \mathcal{N}(x_{t-1}; \mu_\theta(x_t, t), \textstyle{\Sigma}_\theta (x_t, t))
\end{equation}

Diffusion models have been used to solve a wide class of problems in computer vision including semantic segmentation \cite{wang2022semantic}, image and video generation, point cloud generation \cite{luo2021diffusion}, super-resolution \cite{saharia2022image, ho2022cascaded} and anomaly detection \cite{wyatt2022anoddpm}. These models can be either unconditional or conditional, depending on the specific application. Most tasks in computer vision, that are specific to any application, require conditional diffusion models. Depending on the desired outcome, the conditioning can be on class labels, semantic maps, images or even graphs. Diffusion models conditioned on class labels can incorporate desired properties in the generated samples whereas those conditioned on images and semantic maps are able to incorporate rich semantics in the generated samples. 

Diffusion models have been applied to problems related to image super-resolution \cite{chen2017face}, inpainting \cite{yuan2019image} and image-to-image translation \cite{isola2017image}. \cite{saharia2022image, ho2022cascaded} have used diffusion models for generating super-resolution images. \cite{saharia2022palette} has proposed a unified conditional diffusion network for different tasks like inpainting, colorization and JPEG restoration. \cite{wang2022semantic} has shown that semantic representations learned with latent diffusion models can aid semantic segmentation problems and \cite{mei2023vidm, yang2022diffusion} show the implementation of diffusion models in high quality video generation tasks. In \cite{luo2021diffusion}, the authors have proposed a novel method for generating point clouds using diffusion models and \cite{wyatt2022anoddpm} shows how diffusion models can be used for anomaly detection.

Numerous endeavors are exploring diffusion models that exhibit distinct noise structures. \cite{wang2023binary} utilize the binary latent space for representing the images and each image patch is encoded with a latent binary vector. \cite{nachmani2021non} shows that noise models that mix Gaussian and Gamma distributions can generate better results than the Gaussian model alone in certain applications. \cite{jolicoeur2023broadcast} puts forth a framework for a generalized noise model and utilized the Method of Moments for optimizing the often intractable non-Gaussian intermediate steps. In \cite{stevens2023removing}, the authors have use diffusion models for removing structured noise in ill-posed inverse problems having non-Gaussian noise models.

Despite such a variety of works on diffusion models, there has been no work that particularly considers signal-dependent speckle, which is a source of corruption in various practical applications, particularly where coherent imaging is applied.

\section{Diffusion Models for Multiplicative Noise}
\label{noise_model}

This section will formulate diffusion models for images corrupted with multiplicative noise. Formally, an image corrupted with speckle noise can be expressed as:
\begin{equation}
    \label{multiplicative}
    I_t = I_0 + I_0 \, \epsilon_t
\end{equation}

where, $\epsilon_t$ is sampled from a zero-mean Gaussian distribution with variance $\alpha_t^2$, i.e, 
\begin{equation}
    \label{noise_Std}
    \epsilon_t = \mathcal{N}(0, \alpha_t^2)
\end{equation}

Now, for any $\epsilon$ sampled from the standard Gaussian, we have
\begin{equation}
    \epsilon_t = \alpha_t \, \epsilon
\end{equation}

Substituting the expression of $\epsilon_t$ in  (\ref{multiplicative}), 
\begin{equation}
    I_t = I_0 + I_0 \, \alpha_t \, \epsilon
\end{equation}
$$I_t = I_0(1 + \alpha_t \epsilon)$$
The logarithm of the noisy image $I_t$ will be
\begin{equation}
    \log \, I_t = \log \, I_0 + \log \, (1+ \alpha_t \, \epsilon)
\end{equation}

If we have $| \alpha_t \epsilon| < 1$ for all $\alpha_t$, we will have:

\begin{equation}
    \label{multiplicative_final_noise}
    \log \, I_t = \log \, I_0 + \alpha_t \, \epsilon
\end{equation}
where $\epsilon \sim \mathcal{N}(\mathbf{0}, \mathbf{I})$.

In (\ref{multiplicative_final_noise}), $\alpha_t$ represents the noise schedule. Following a linear noise schedule, for any $1 \, \le \, t \, \leq \, T$, let
\begin{equation}
    \label{delta}
    \delta = \alpha_t - \alpha_{t-1}
\end{equation}

Thus, (\ref{multiplicative_final_noise}) can be written as 

$$\log \, I_t = \log \, I_0 + \alpha_t \, \epsilon$$ 
$$\text{or, } \log \, I_t = \log \, I_0 + (\alpha_{t-1} + \delta) \, \epsilon$$ 
$$\text{or, } \log \, I_t = \log \, I_0 + \alpha_{t-1} \, \epsilon + \delta \, \epsilon$$ 

Hence, we have
\begin{equation}
    \label{multiplicative_single_step_forward}
    \log \, I_t = \log \, I_{t-1} + \delta \, \epsilon
\end{equation}

In this sense, the current diffusion approach is somewhat of a homomorphic technique.  In homomorphic filtering, the image intensities are transformed by a logarithmic operation which yields an additive relationship between signal and noise from the multiplicative case.

\subsection{Forward Process}
We can represent the forward processes as follows:

\begin{equation}
    \label{mul_q_t_1}
    q(\log \, I_t| \log \, I_{t-1}) \sim \mathcal{N}(\log \, I_{t}; \log \, I
_{t-1}, \delta^2 \, \mathbf{I})
\end{equation}

and
\begin{equation}
\label{mul_q_0}
    q(\log \, I_t| \log \, I_0) \sim \mathcal{N}(\log \, I_t; \log \, I_0, \alpha_t^2 \, \mathbf{I})
\end{equation}
\\
Using (\ref{mul_q_t_1}) and (\ref{mul_q_0}) and applying Bayes' rule,\\ \\
$q(\log I_{t-1} | \log I_t, \log I_0) \\ \\
= \dfrac{q(\log I_{t} | \log I_{t-1}, \log I_0) \, q(\log I_{t-1} | \log I_0)}{q(\log I_{t} | \log I_0)}$ \\ \\ 
= $\dfrac{\mathcal{N}(\log I_{t} ; \log I_{t-1}, \delta^2 \mathbf{I}) \, \mathcal{N}(\log I_{t-1} ; \log I_{0}, \alpha_{t-1}^2 \mathbf{I})}{\mathcal{N}(\log I_{t} ; \log I_{0}, \alpha_{t}^2 \mathbf{I})}$ \\ \\
= $\exp \{-\dfrac{1}{2} [ \dfrac{(\log I_t - \log I_{t-1})^2}{\delta^2} +  \dfrac{(\log I_{t-1} - \log I_{0})^2}{\alpha_{t-1}^2} - \dfrac{(\log I_{t} - \log I_{0})^2}{\alpha_{t}^2}]\}$ \\ \\
= $\exp \{ -\dfrac{1}{2} [ \log I_{t-1}^2 \left(\dfrac{1}{\delta^2} + \dfrac{1}{\alpha_{t-1}^2}\right) - 2 \log I_{t-1} \left(\dfrac{\log I_t}{\delta^2} + \dfrac{\log I_0}{\alpha_{t-1}^2}\right) + C(\log I_t, \log I_0)] \}$; (where $C(\log I_t, \log I_0)$ is a constant involving $I_t$ and $I_0$) \\ \\
$\propto \exp \{-\dfrac{1}{2}[\log I_{t-1}^2 \left(\dfrac{\alpha_{t-1}^2 + \delta^2}{\alpha_{t-1}^2 \, \delta^2}\right) - 2 \log I_{t-1} \left( \dfrac{\alpha_{t-1}^2 \log I_t + \delta^2 \log I_0}{\alpha_{t-1}^2 \, \delta^2}\right)
]\}$ \ \\ \\
= $ \exp \{-\dfrac{1}{2}\dfrac{1}{\left(\dfrac{\alpha_{t-1}^2 \delta^2}{\alpha_{t-1}^2 + \delta^2}\right)}[\log I_{t-1}^2 - 2 \log I_{t-1}\left( \dfrac{\alpha_{t-1}^2 \log I_t + \delta^2 \log I_0}{\alpha_{t-1}^2 \ + \delta^2}\right)]  \}$
\\ \\
$q(\log \, I_{t-1} | \log \, I_t) \sim \mathcal{N}(\log \, I_{t-1}; \mu_q(\log \, I_t, \log \, I_0 ), \textstyle\Sigma_q(t)\mathbf{I})$ where $$\Sigma_q(t) = \dfrac{\alpha_{t-1}^2 \, \delta^2}{\alpha_{t-1}^2 + \delta^2}$$ and
$$\mu_q (\log \, I_t, \log \, I_0) = \dfrac{\alpha_{t-1}^2 \, \log \, I_t + \delta^2 \, \log \, I_0}{\alpha_{t-1}^2 + \delta^2}$$ 

\subsection{Reverse Process}

The reverse process can be formulated as:
\begin{equation}
    p(\log \, I_{0:T}) = \prod_{t=1}^T p_\theta(\log \, I_{t-1}|\log \, I_t)
\end{equation}

Unlike DDPM \cite{ho2020denoising}, 
$$p(\log I_T) \neq \mathcal{N}(\log \, I_T; \mathbf{0}, \mathbf{I})$$ This is because SDDPM considers multiplicative noise and the constituent image regions are corrupted to different extents. Since the forward process $q(\log \, I_{t-1} | \log \, I_t)$, we can set the reverse process to be Gaussian as well. Moreover, since $$\Sigma_q(t) = \dfrac{\alpha_{t-1}^2 \, \delta^2}{\alpha_{t-1}^2 + \delta^2}$$ the model estimates the mean of the Gaussian ($\mu_{\theta}$) and the variance is kept as $\Sigma_q(t)$ \cite{ho2020denoising}. \\

The reverse process $p_\theta$ is: \\
 
$p_\theta (\log \, I_{t-1}| \log \, I_t) = \\ \\
\mathcal{N}(\log \, I_{t-1}; \mu_\theta(\log \, I_t, \tau), \textstyle\Sigma_\theta(t)\mathbf{I})$ \\

for $2 \le t \le T$ and $\tau = \{1, 2, ... T\}$ represents the noise schedule. Following \cite{ho2020denoising}, we set $\textstyle \Sigma_\theta(t) = \textstyle \Sigma_q(t)$ as the individual variances corresponding to different timesteps are independent of the images $I_t$. The mean for $\mu_\theta(\log \, I_t, \tau )$ will be: 
$$ \dfrac{\alpha_{t-1}^2 \, \log \, I_t + \delta^2 \, \log \, {f_\theta}(I_t, \tau)}{\alpha_{t-1}^2 + \delta^2}$$ \\

where ${f_\theta}(I_t, \tau)$ is the prediction of the network for an input image corrupted with noise $\alpha_t$.

\section{Training Objective}

The forward process $q(\log \, I_{t-1}| \log \, I_t)$ and the reverse process $p_\theta(\log \, I_{t-1}| \log \, I_t)$ are both Gaussian, and the variances are same, i.e, $$\textstyle \Sigma_\theta(t) = \textstyle \Sigma_q(t)$$
The Kullback–Leibler (KL) divergence ($D_{KL}$) between the forward and the reverse processes can be minimized as \cite{luo2022understanding}:
$$\textstyle{\underset{\theta}{\mathrm{argmin}}} \,  D_{KL} (q (\log \, I_{t-1}| \log \, I_t, \log \, I_0)) || p_{\theta} (\log \, I_{t-1}| \log \, I_t))$$
\\
= $\textstyle{\underset{\theta}{\mathrm{argmin}}} \, D_{KL} (\mathcal{N}(\log \, I_{t-1}; \mu_q(\log \, I_t, \log \, I_0 ), \textstyle\Sigma_q(t)\mathbf{I}) \, || \\  \mathcal{N}(\log \, I_{t-1}; \mu_\theta(\log \, I_t, \tau), \textstyle\Sigma_q(t)\mathbf{I}))$. 
\\ \\
The above equation can be further simplified as:
\begin{equation}
    \label{mul_loss}
    \textstyle{\underset{\theta}{\mathrm{argmin}}} \dfrac{1}{2 \textstyle \Sigma_q(t)} \left [ || \mu_\theta - \mu_q||_2^2\right]
\end{equation}
where, 
$$\mu_q = \mu_q(\log \, I_t, \log \, I_0 )$$
and
$$\mu_{\theta} = \mu_\theta(\log \, I_t, \log \, {f_\theta}(I_t, \tau))$$
Substituting the expressions of $\mu_q (\log \, I_t, \log \, I_0) $ and $\mu_\theta(\log \, I_t, \tau)$ in  (\ref{mul_loss}), the loss function reduces to:
\begin{equation}
    \label{data_consistency}
    \mathcal{L}_{D} \,  = \mathbf{E}\left[ ||{f_\theta}(I_t, \tau) -  I_0||_2^2\right]
\end{equation}

To summarize, we train a neural network model $p_\theta$ which learns to remove the multiplicative noise affecting the original images. Given any image, it learns to approximate $\mu_\theta$ such that it matches $\mu_q$ as $q(\log \, I_{t-1}| \log \, I_t)$ and $p_\theta(\log \, I_{t-1}| \log \, I_t)$ are both Gaussian distributions with the same variance but different means.
\\
\begin{algorithm}
    \label{Training_algo}
    \caption{Training Algorithm}
    \hspace*{\algorithmicindent} \textbf{Input}: \{$I_0$\}, $\tau$, $f_\theta$\\
    \hspace*{\algorithmicindent} \textbf{Output}: trained $f_\theta$\\
  \begin{algorithmic}[1]
  
    \WHILE{not converged}
    \STATE $I_0 \sim q(I_0)$
    \STATE $t \sim Uniform({1, ..., T})$
    \STATE $\epsilon \sim \mathcal{N}(\mathbf{0}, \mathbf{I})$
    \STATE $\log \, I_t = \log \, I_0 + \alpha_t \, \epsilon$
    \STATE Compute gradient descent on 
    $\nabla_\theta ||f_\theta(I_t, \tau) - I_0||_2^2$
    \STATE Update $\theta$
    \ENDWHILE
  \end{algorithmic}
\end{algorithm}

Algorithm 1 describes the training algorithm. At any training epoch, the input image $I_0$ is corrupted with noise following a chosen noise schedule $\alpha_t$ where $t \in [1, T]$. The noise level introduced in the input image is determined by $t$. The noisy image $I_t$ is given to the neural network $f_{\theta}$ and the network predicts the original input image. For the denoising task, the trained model takes as input a noisy image and predicts the denoised image.  

\section{Experiments}

\subsection{Datasets}

We have performed our experiments on the UC Merced Land Use dataset \cite{yang2010bag} and ultrasound images \cite{zhang2022ultrasound, WinNT}. The land use dataset contains images from 21 different categories and each category has 100 images. Each image is 256 x 256 in size and they are extracted from larger images obtained from different urban areas. The ultrasound images used in this paper comprise two kinds. The first set of ultrasound images are those of the common carotid artery (CCA) collected from 10 volunteers. The images sizes vary between $230 \times 390$ and $450 \times 600$. The second set of ultrasound images comprises images of the fetal head, lymph nodes and the brachial plexus \cite{zhang2022ultrasound}, where all images are $256 \times 320$ in size. 
\\
\begin{figure}[!htb]
    \centering
    \includegraphics[scale = 0.5]{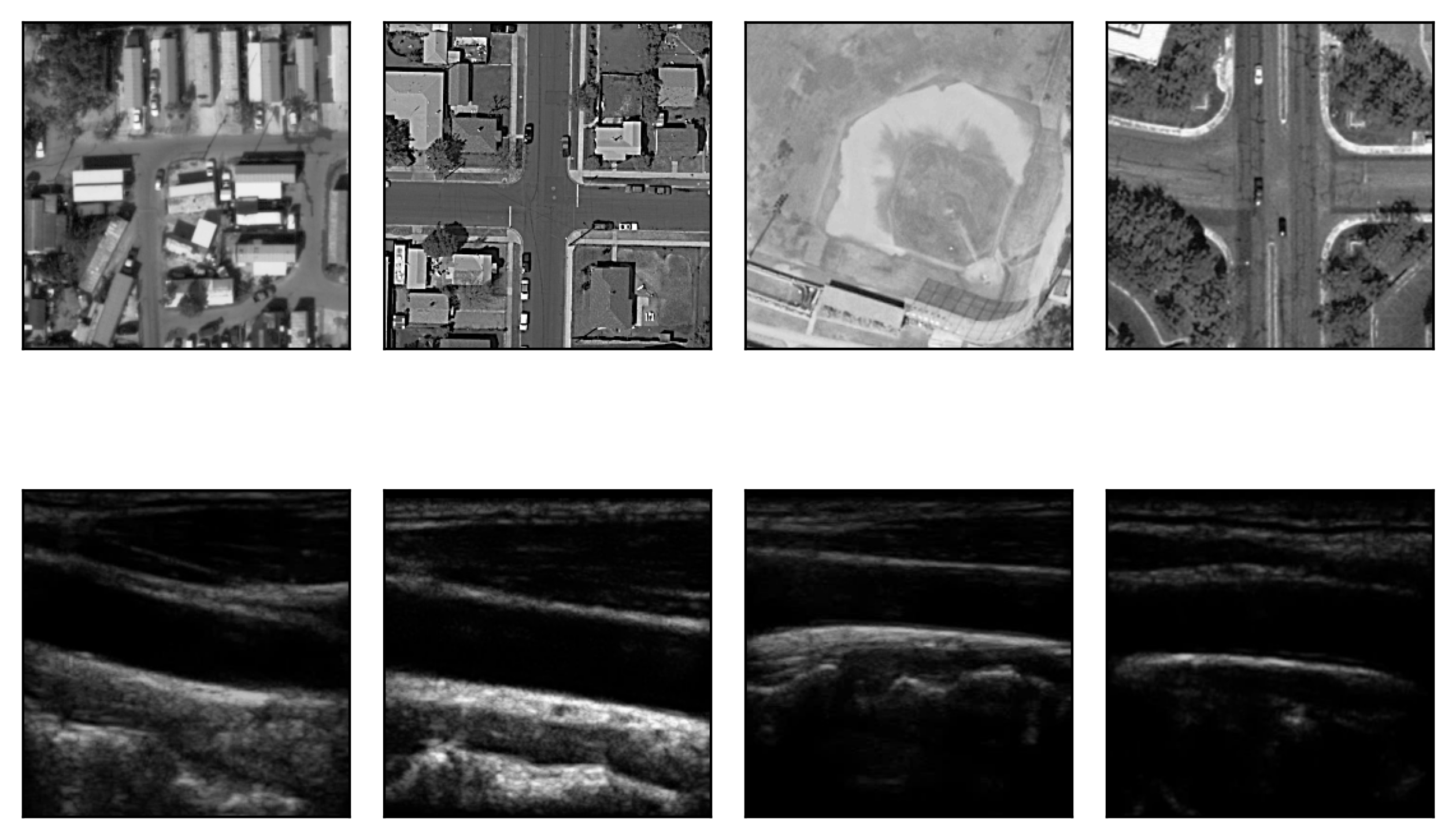}
    \caption{Sample images from the datasets used for the experiments. The top row shows images from the land-use dataset and the bottom row shows images from the ultrasound dataset.}
    \label{fig:enter-label}
\end{figure}

\begin{table*}[!htb]
\caption{Results on Land-use Dataset}
\centering
\label{tab_SAR}
 \renewcommand{\arraystretch}{1.2}
\begin{tabular}
{| p{0.098\linewidth} | p{0.06\linewidth} | p{0.06\linewidth} | p{0.06\linewidth} | p{0.06\linewidth} | p{0.06\linewidth} | p{0.06\linewidth} | p{0.06\linewidth} | p{0.06\linewidth} | }
\hline
\multirow{2}{*}{Method} & \multicolumn{2}{c |}{$\alpha_t = 0.0771$} & \multicolumn{2}{c |}{$\alpha_t = 0.2015$} & \multicolumn{2}{c |}{$\alpha_t = 0.3756$} &\multicolumn{2}{c |}{$\alpha_t = 0.5$} \\ \cline{2-9}

 & PSNR & SSIM & PSNR & SSIM & PSNR & SSIM & PSNR & SSIM\\
 \hline
 \hline
SRAD \cite{yu2002speckle} & 25.52 & 0.855 & 24.96 & \textbf{0.800} & 22.14 & 0.639 & 19.23 & 0.489 \\
 \hline
 BM3D \cite{dabov2007image} & 21.30 & 0.777 & 18.45 & 0.712 & 10.39 & 0.296 & 9.29 & 0.177 \\
\hline
NLMeans \cite{buades2005non} & 21.21 & 0.846 & 12.80 & 0.511 & 9.59 & 0.227 & 8.79 & 0.150 \\
\hline
\hline
DnCNN \cite{zhang2017beyond} & 22.34 & 0.699 & 22.09 & 0.632 & 21.76 & 0.544 & 21.96 & 0.489 \\
\hline
SCU \cite{zhang2022practical} & 24.13 & 0.755 & 24.32 & 0.692 & 24.26 & 0.572 & 24.72 & 0.479 \\
\hline
SwinIR \cite{liang2021swinir} & 23.50 & 0.772 & 23.50 & 0.688 & 23.06 & 0.549 & 22.89 & 0.452 \\
\hline 
SDDPM (Ours) & \textbf{29.97} & \textbf{0.876} & \textbf{27.79} & 0.771 & \textbf{26.04} & \textbf{0.709} & \textbf{25.00} & \textbf{0.662} \\
\hline

\end{tabular}
\end{table*}
 \renewcommand{\arraystretch}{1}

\begin{table*}[htb]
\caption{Results on Ultrasound Datasets}
\centering
\label{tab_US}
\renewcommand{\arraystretch}{1.3}
\begin{tabular}
{| p{0.098\linewidth} | p{0.06\linewidth} | p{0.06\linewidth} | p{0.06\linewidth} | p{0.06\linewidth} | p{0.06\linewidth} | p{0.06\linewidth} | p{0.06\linewidth} | p{0.06\linewidth} | }
\hline
\multirow{2}{*}{Method} & \multicolumn{2}{c |}{$\alpha_t = 0.0771$} & \multicolumn{2}{c |}{$\alpha_t = 0.2015$} & \multicolumn{2}{c |}{$\alpha_t = 0.3756$} &\multicolumn{2}{c |}{$\alpha_t = 0.5$} \\ \cline{2-9}

 & PSNR & SSIM & PSNR & SSIM & PSNR & SSIM & PSNR & SSIM\\
 \hline
 \hline
SRAD \cite{yu2002speckle} & 29.21 & 0.857 & 28.78 & 0.854 & 26.63 & 0.836 & 23.95 & 0.780 \\
 \hline
 BM3D \cite{dabov2007image} & 25.87 & 0.758 & 26.66 & 0.746 & 19.74 & 0.560 & 17.88 & 0.461 \\
\hline
NLMeans \cite{buades2005non} & 31.86 & \textbf{0.922} & 24.84 & 0.773 & 18.89 & 0.534 & 17.21 & 0.405 \\
\hline
\hline
DnCNN \cite{zhang2017beyond} & 17.79 & 0.281 & 17.93 & 0.277 & 17.99 & 0.263 & 18.13 & 0.254 \\
\hline
SCU \cite{zhang2022practical} & 25.33 & 0.854 & 25.47 & 0.840 & 25.47 & 0.802 & 24.96 & 0.780 \\
\hline
SwinIR \cite{liang2021swinir} & 21.33 & 0.683 & 21.11 & 0.665 & 20.25 & 0.644 & 19.48 & 0.599 \\
\hline
SDDPM (Ours) &\textbf{ 32.81} & 0.895 & \textbf{31.71} & \textbf{0.883} & \textbf{29.95} & \textbf{0.854} & \textbf{28.72} & \textbf{0.828} \\
\hline

\end{tabular}
\end{table*}

\begin{figure*}[!htb]
\begin{center}
\includegraphics[width=0.9\textwidth]{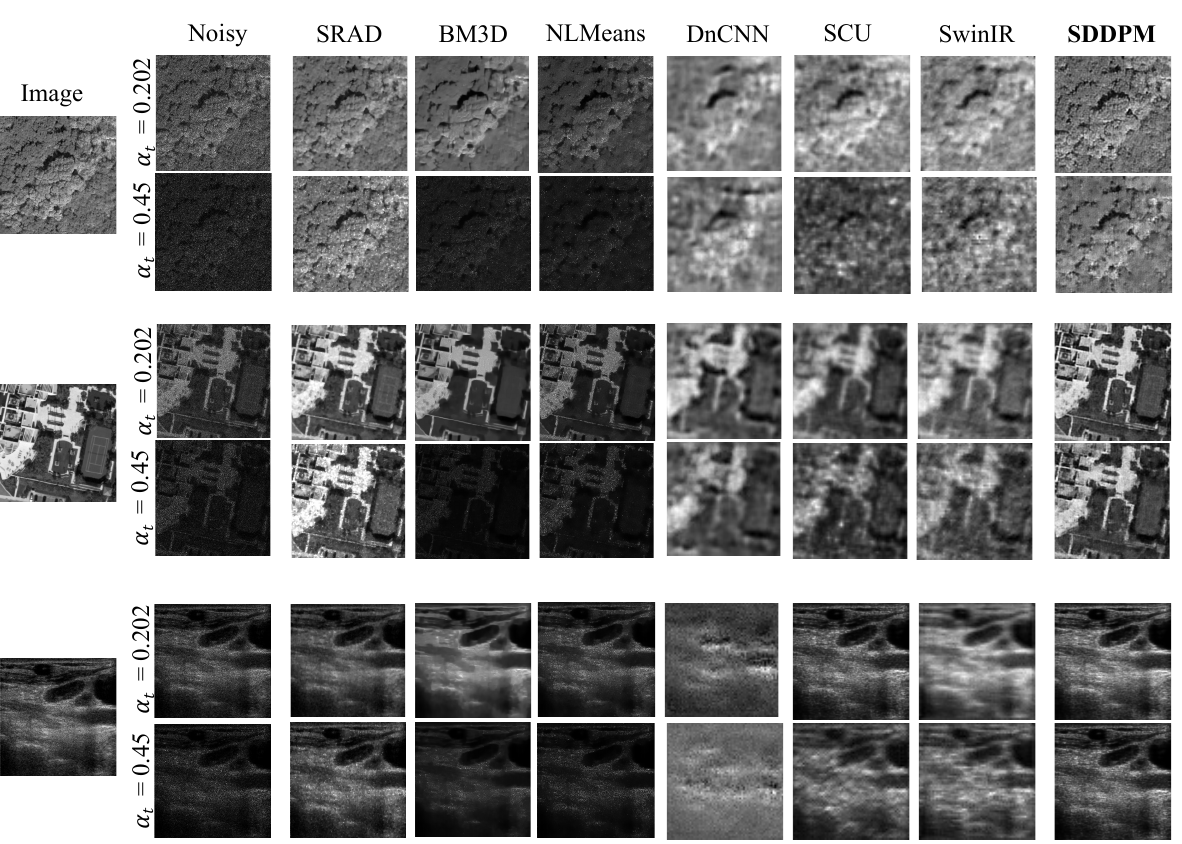}
\end{center}
\caption{Results generated by the proposed model for different noise levels. The first column shows the original image. All images in a row except the first one show noisy and denoised versions of the original image. }
\label{comparison_imgs}
\end{figure*}

\begin{figure*}[!htb]
\begin{center}
\includegraphics[width=0.9\textwidth]{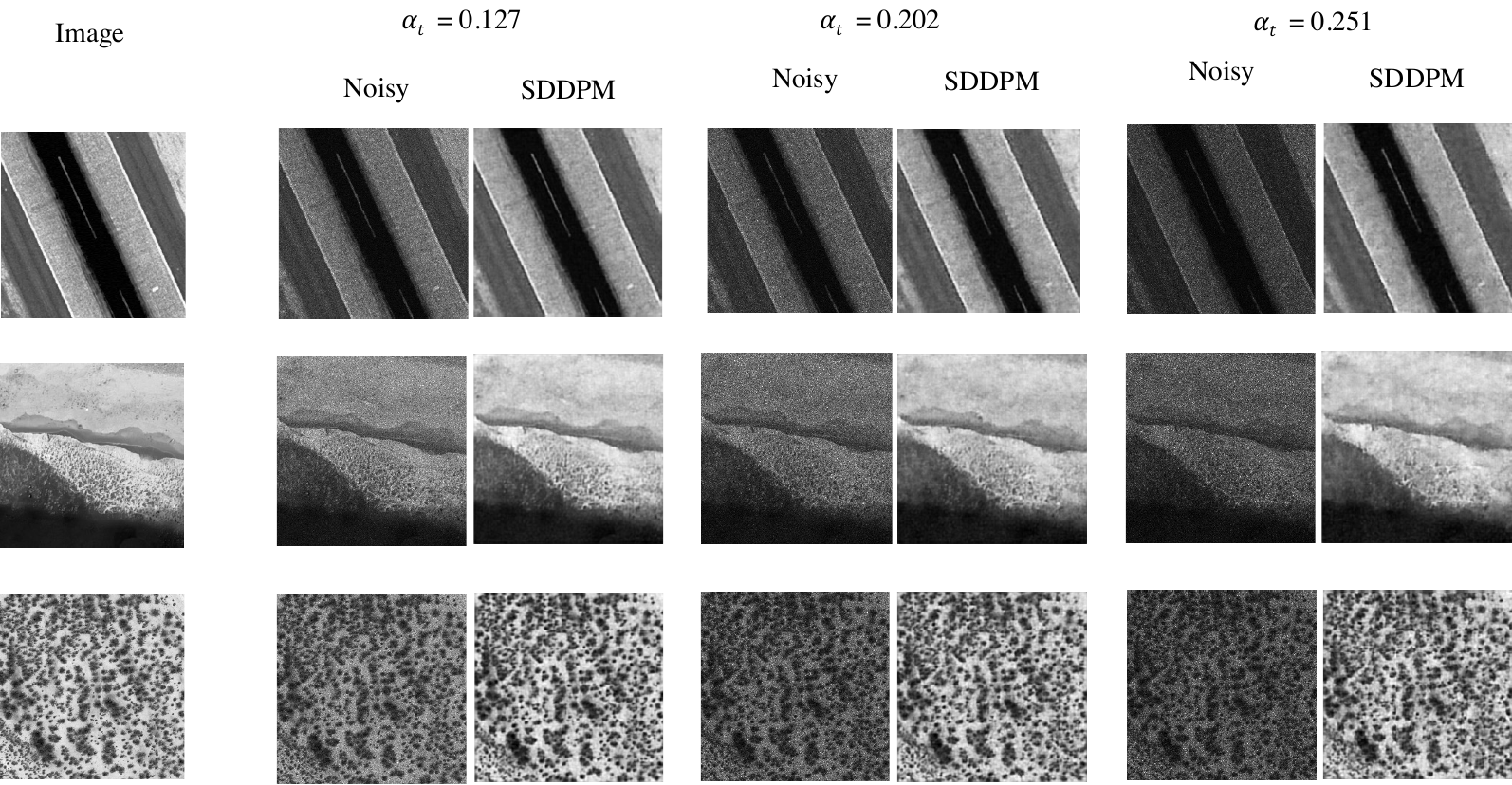}
\end{center}
\caption{Results generated by the proposed model for different noise levels. The first column shows the original image. All images in a row except the first one show noisy and denoised versions of the original image. }
\label{sar_lines_texture}
\end{figure*}
\begin{figure*}[!htb]
\begin{center}
\includegraphics[width=0.9\textwidth]{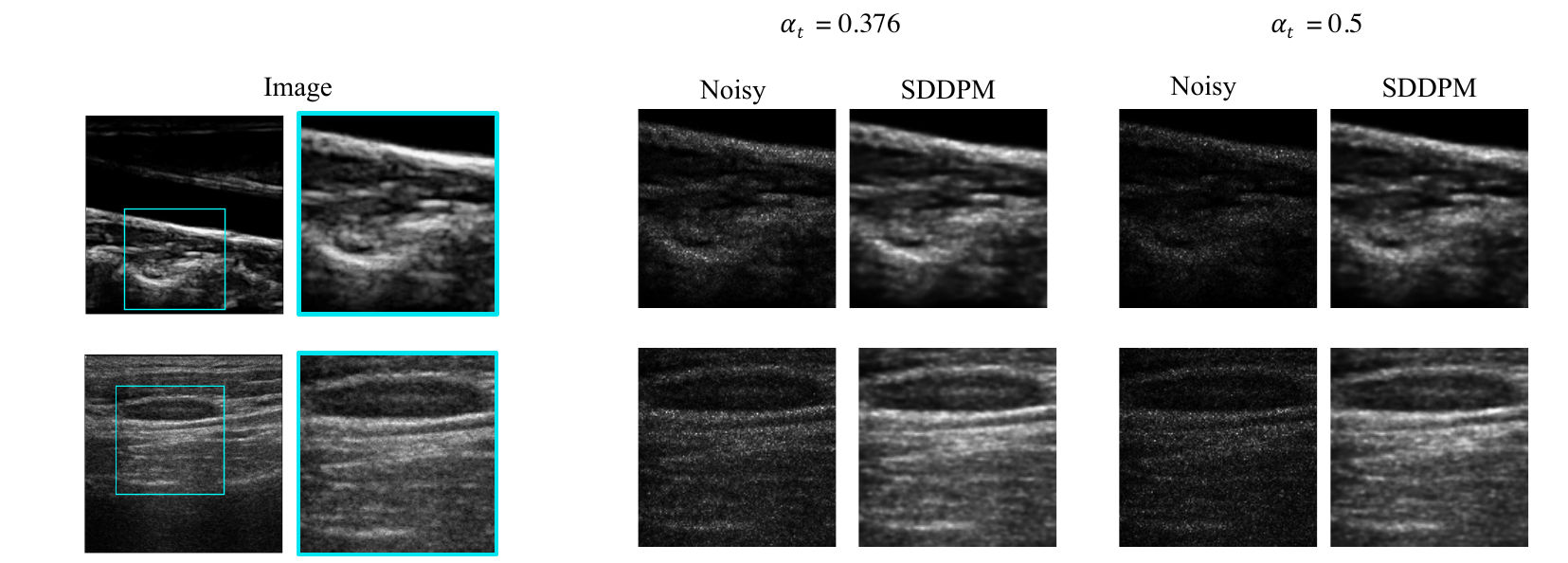}
\end{center}
\caption{Column 1: Original image with a particular selected section shown in the colored box. Column 2: The enlarged view of the area inside the box. Columns 3-6 show noisy and denoised versions of Column 2 for different noise levels specified by $\alpha_t$.}
\label{US_denoised}
\end{figure*}
\subsection{Baseline Models}
\label{baselines}

We have compared our model to some classical methods and some recent deep neural network models. We have selected an array of denoising methods, encompassing those specially tailored for speckle reduction and others with a broader scope. In terms of comparisons to neural network models, we selected systems in which the input is solely the noisy image and the training methodology is similar to ours. All together, we compare to three classical approaches and to three recent solutions based on deep neural networks.
\\
\subsubsection{Classical Methods}

\begin{itemize}

    \item  The block-matching and 3D filtering (BM3D) was proposed for image denoising in \cite{dabov2007image}. A noisy image is partitioned into overlapping blocks and similar blocks are identified and grouped together. Collaborative filtering is applied on similar patches in a group to estimate the underlying clean signal. BM3D achieves state-of-the-art denoising performance by leveraging the redundancy and consistency present in groups of similar patches, without requiring prior knowledge about specific noise characteristics. 
    \item The non-local means \cite{buades2005non} (NLMeans) denoising algorithm preserves the high frequency details while effectively reducing noise in images. For a patch centered around a pixel, NLMeans identifies similar patches centered around other pixels and the patches are assigned weights based on their similarity to the central patch. Patches that are more similar to the central patch receive higher weights. Each noisy pixel is replaced by the average pixel value of the similar patches. This distinctive approach captures non-local information, allowing it to excel at maintaining fine details and global image structure. NLMeans is adaptable to various noise types and levels and is particularly adept at noise reduction in textured areas. 
    \item SRAD \cite{yu2002speckle} is a partial differential equation method that is particularly configured for speckled images and considers the local structure of the image. The core idea of SRAD is to iteratively adjust pixel intensities based on the local variations in intensity. This approach effectively smooths regions with homogenous intensity values while retaining image features across the edges. The diffusion equation in SRAD incorporates an anisotropic term that allows diffusion to be adjusted based on the local coefficient of variation rather than the gradient magnitude. \\

\end{itemize}

\subsubsection{Deep Neural Network Models}

\begin{itemize}
    \item In \cite{zhang2017beyond}, a feed-forward convolutional neural network-based image denoising algorithm has been proposed. Many image denoising methods involve training specific models for additive Gaussian noise for a specific noise level. In contrast, this paper proposed a method for blind Gaussian denoising. Instead of an explicit image prior, the proposed model learns to predict noise from a given noisy image.
    \item SwinIR \cite{liang2021swinir} is a transformer-based image restoration model that involves three phases, namely, shallow feature extraction, deep feature extraction and image reconstruction. The shallow features are extracted by a convolutional layer whereas the deep features are extracted by a sequence of residual swin transformer blocks (RSTB) followed by a final convolutional layer. Finally, the image reconstruction layer is implemented with a sub-pixel convolutional layer \cite{caballero2017real}. Combining the shallow and deep features helps the model capture both the low frequency and high frequency details present in the input images. The swin transformer layer \cite{liu2021swin} differs from the original transformer layer \cite{vaswani2017attention} in the way it calculates local attention using a shifted window mechanism. 
    \item A deep blind image denoising multi-scale UNet is proposed in \cite{zhang2022practical} where modeling of local and non-local features is boosted by swin-conv blocks. The Swin-Conv-UNet(SCU) has residual convolution blocks and swin transformer blocks \cite{liu2021swin} and can thereby achieve local and non-local modeling of the input data. The authors have also proposed a noise synthesis model that improves the performance of the proposed model for different images. 


\end{itemize}

\subsection{Evaluation Metrics}

We have used peak signal-to-noise ratio (PSNR) and the structural similarity index (SSIM) to evaluate the quality of the denoised images. If $I_{max}$ denotes the maximum intensity of an image, the PSNR and SSIM metrics are defined as follows.

\begin{equation}
    PSNR = 10 \, \log_{10} \left(\dfrac{I_{max}^2}{|I_0 - \hat{I_0}|^2}\right)
\end{equation}

\begin{equation}
    \label{ssim}
    SSIM(I_0, \hat{I_0}) = \dfrac{(2 \mu_{I_0} \mu_{\hat{I_0}} + c_1)(2\sigma_{I_0 \, \hat{I_0}} + c_2)}{(\mu_{I_0}^2 + \mu_{\hat{I_0}}^2 + c_1)(\sigma_{I_0}^2 + \sigma_{\hat{I_0}}^2 + c_2)}
\end{equation}
    
High PSNR indicates that the denoised image resembles the original image closely and retains the high frequency details present in the original image. In (\ref{ssim}), $\mu_{I_0}$ and $\mu_{\hat{I_0}}$ are the mean pixel values of $I_0$ and $\hat{I_0}$ respectively, $\sigma_{I_0}^2$ and $\sigma_{\hat{I_0}}^2$ are the variances, $\sigma_{I_0 \, \hat{I_0}}$ is the covariance and $c_1$ and $c_2$ are constants. SSIM is a perceptual metric that is sensitive to the local and global variations in the images. For comparing the similarity between two images, SSIM considers the structural information along with intensity and contrast. SSIM values range between -1 and 1, where higher values indicate higher fidelity. 

\subsection{Experimental Setup}

Both the datasets comprise multiple categories and the number of images in each of these categories is limited. We split the dataset into train, validation and test sets randomly and augmented the training images. We applied random rotations to each image in the training set and added Gaussian noise with mean $\mu = \{0, 0.05\}$ and variance $\sigma^2 = \{0, 0.001\}$ to create the final training dataset. The models were trained for 100 epochs. The initial learning rate was set as $0.05$ and was reduced by a factor of $10$ after every $20$ epochs. The stochastic gradient descent (SGD) algorithm was used for optimizing all the models. For all the experiments, $\alpha_t$ was varied between $[ 0.005, 0.5 ]$ and the timesteps $T$ was set as $200$. All models were trained on $64 \times 64$ randomly cropped patches and the denoising step was performed on $128 \times 128$ images.

\section{Results}
Fig. \ref{comparison_imgs} shows the qualitative results obtained by the different denoising algorithms for different noise levels. We have shown the results for two images from the UC Merced Land Use dataset and one ultrasound image. We have shown the results for $\alpha_t = 0.202$ and $\alpha_t = 0.45$ for every image. It should be noted that $\alpha_t = 0.45$ corresponds to very high noise and the images are corrupted significantly. Column 2 shows the noisy images generated for each of the different cases. Columns 3 - 8 show the denoised images generated by each of the different baseline algorithms. Column 9 shows the results generated by the proposed model SDDPM. It should be noted that the images generated by SRAD \cite{yu2002speckle} and SDDPM closely resemble the original image even for $\alpha_t = 0.45$. However, under such high noise conditions, the denoised images obtained with SRAD algorithm tend to have some artifacts which are not present in the denoised images obtained with SDDPM. Most of the high frequency information present in the original image is recovered by SDDPM even under extreme noise conditions. 

In Fig. \ref{sar_lines_texture} and \ref{US_denoised} we take a closer look at some more denoised images generated by SDDPM. For both natural images and ultrasound images, the denoising algorithm should be able to reconstruct the details present in the original images as accurately as possible. Most denoising algorithms often suffer from excessive smoothing across the edges. There is often a trade-off between noise removal and edge preservation. This is also evident from the different results shown in Fig. \ref{comparison_imgs}. In Fig. \ref{sar_lines_texture}, it can be seen that the denoised images generated by SDDPM for different noise levels have very prominent edges and textures. The results show that even for higher noise levels, the images denoised by SDDPM retain most of the structures that are present in the original images. In Fig. \ref{US_denoised}, we have shown the denoised ultrasound images for different noise levels. When medical images are denoised, the intrinsic structures present in the original image need to be retained so that an accurate diagnosis can be made from the denoised images. For denoising algorithms where the denoised images are smoothed out excessively, the important structures originally present in the ultrasound images can be lost. In Fig. \ref{US_denoised}, we have shown an enlarged version of a certain section in each of the images where a lot of details are present.  It can be seen that even for very high noise levels, the denoised images can retrieve most of the information present in the original image. Unlike other denoising algorithms, the images denoised with SDDPM can reconstruct the details present in the original image while maintaining the intensity variations at the edges. 

Tables \ref{tab_SAR} and \ref{tab_US} show the numerical scores of PSNR and SSIM metrics used for evaluating the denoised images generated by the different methods for 4 different noise levels. $\alpha_t$ is the standard deviation of noise as shown in (\ref{noise_Std}). We have shown the numerical scores for different $\alpha_t$, where higher $\alpha_t$ indicates more noise. In both tables, we have compared our method to all the baselines mentioned in \ref{baselines}. Linear noise schedule was considered for all experiments and $\alpha_t$ was varied from $0.005$ to $0.5$ in $200$ timesteps. 

\section{Discussions}

\subsection{How does the performance of different models vary under varied noise conditions?}

The performance of SRAD, NLMeans and BM3D does not depend on the maximum noise level present in the training images. This independence is due to the fact that the methods are not data driven and to the fact that each noisy image is processed separately. However, it is worthwhile to see the effect of the maximum noise level (maximum value of $\alpha_t$) on the performance of DnCNN, SCU, SwinIR and SDDPM. We can see from Fig. \ref{sar_lines_texture} and \ref{US_denoised} that SDDPM performs well even in the presence of high noise. Intuitively, any trained model should perform better when the noise is low and the performance drops as the noise is increased. This is what is seen for the SSIM and PSNR scores for SDDPM. However, this is not true for all the baseline models. Tables \ref{tab_SAR} and \ref{tab_US} show that the PSNR and the SSIM scores for most of the baseline models are much lower than the proposed model in all cases. In some cases (DnCNN and SCU), the PSNR scores are somewhat erratic. In these cases, the usual trend is not observed, and the performance metrics vary arbitrarily within a small interval. To further explore this matter, we narrowed down the range of $\alpha_t$ values and conducted experiments for the two datasets.

\begin{figure}[!htb]
\begin{center}
\includegraphics[width=0.45\textwidth]{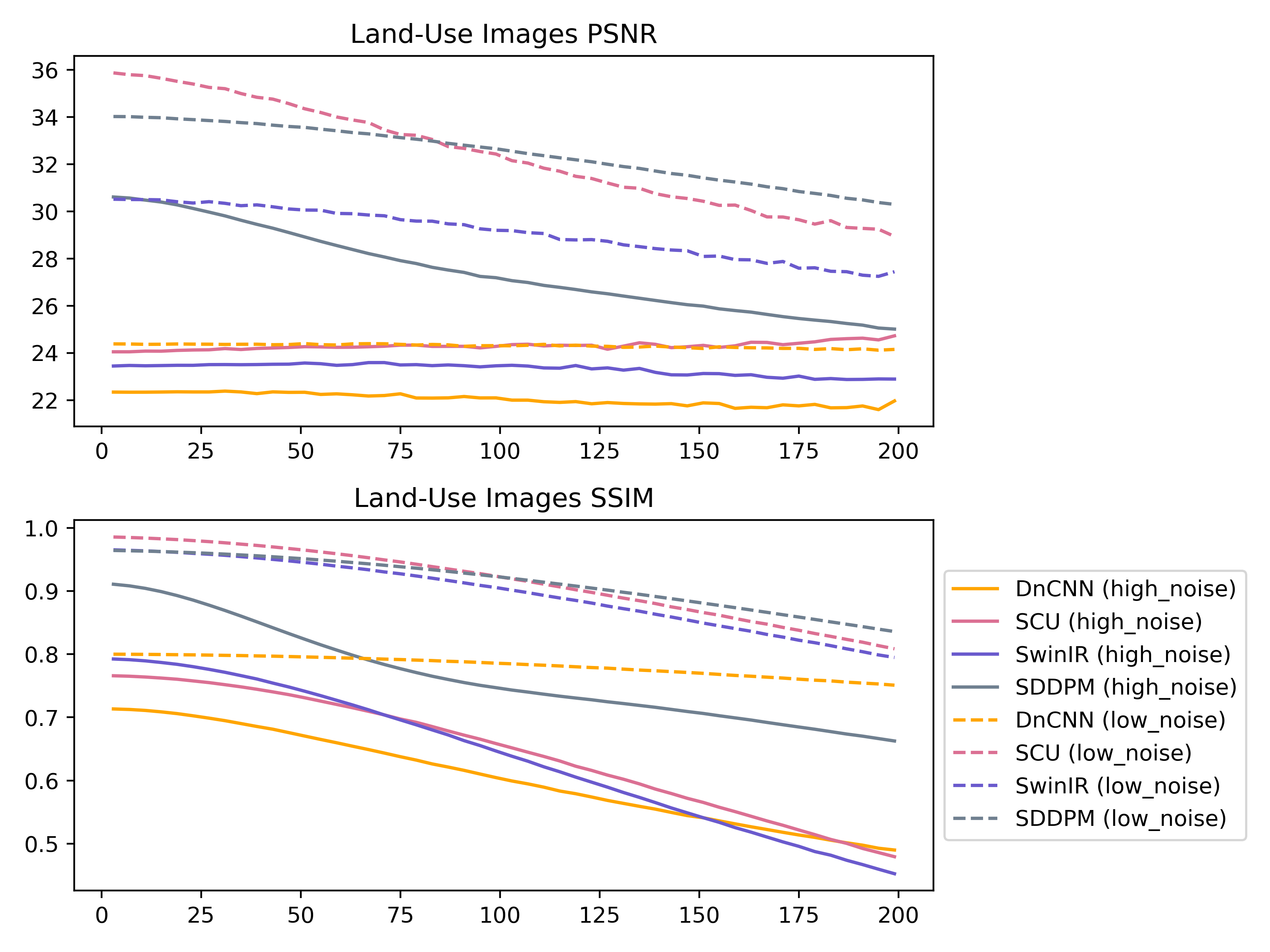}
\end{center}
\caption{PSNR scores (top row) and SSIM scores (bottom row) achieved by DnCNN, SCU, SwinIR and SDDPM for the land-use dataset.}
\label{noise_level_change_SAR}
\end{figure}

\begin{figure}[!htb]
\begin{center}
\includegraphics[width=0.45\textwidth]{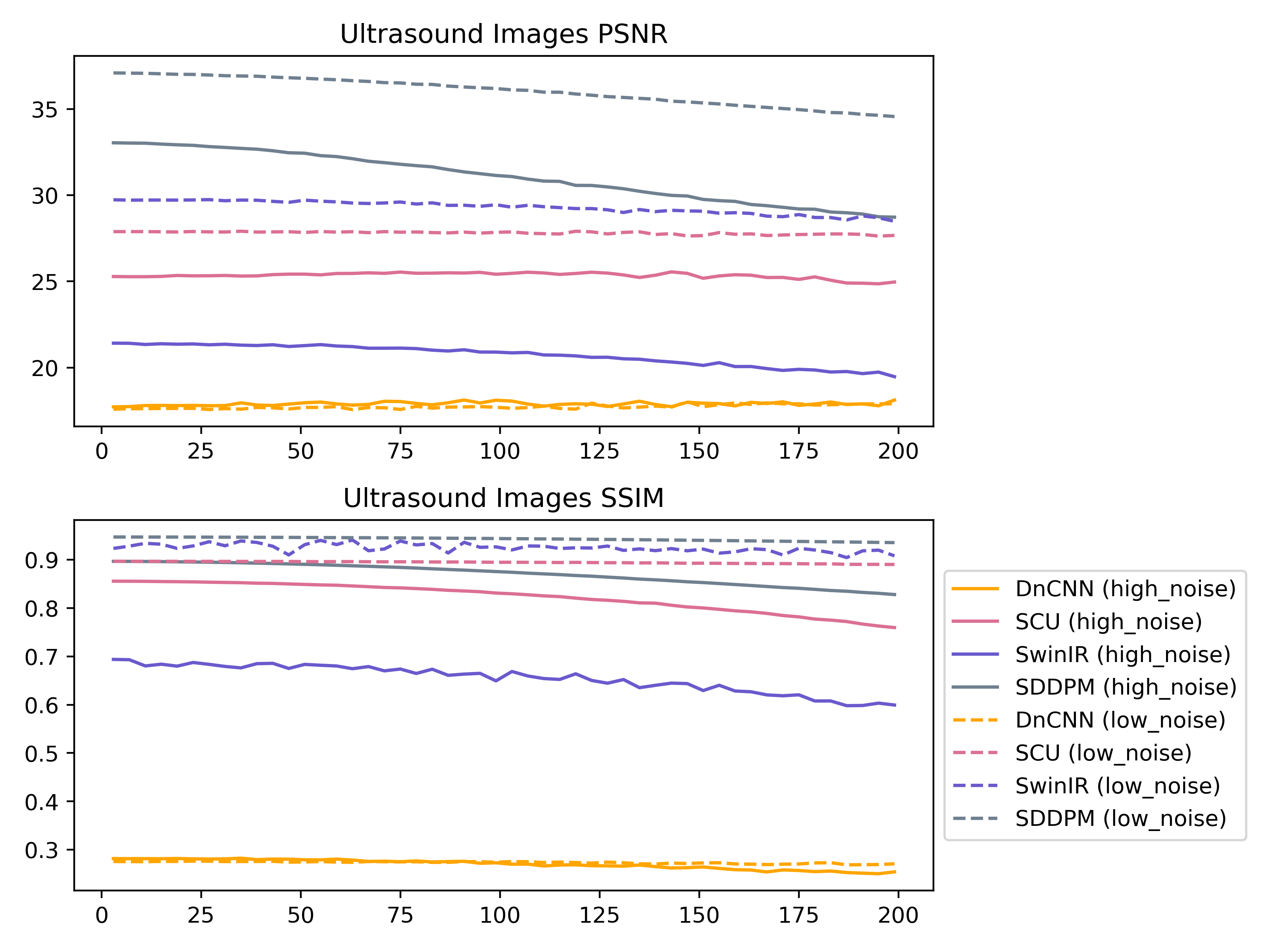}
\end{center}
\caption{PSNR scores (top row) and SSIM scores (bottom row) achieved by DnCNN, SCU, SwinIR and SDDPM for the ultrasound dataset.}
\label{noise_level_change_US}
\end{figure}

In Fig. \ref{noise_level_change_SAR} and \ref{noise_level_change_US}, we have shown how the range of $\alpha_t$ affects the performance of DnCNN, SCU, SwinIR and SDDPM. All the results shown in tables \ref{tab_SAR} and \ref{tab_US} are obtained with $\alpha = [0.005, 0.5]$ and the timesteps $T = 200$. Next, we performed all the experiments with reduced noise where $\alpha = [0.005, 0.1]$ and the timesteps $T = 200$. Low noise refers to $\alpha = [0.005, 0.1]$ whereas high noise refers to $\alpha = [0.005, 0.5]$. When $\alpha_t$ is low, we see that PSNR and SSIM are higher and the values decrease as $\alpha_t$ increases. It should be noted that the SSIM scores for most of the models are high both both the datasets when $\alpha = [0.005, 0.1]$, i.e, when the noise is low. The scores change significantly when the overall noise level is increased ($\alpha = [0.005, 0.5]$). This is because during training, we are adding different levels of noise and the baseline convolutional architectures are more capable of denoising when the noise levels do not vary too much. 

The proposed method SDDPM performs better than the baseline models with different noise levels. When $\alpha = [0.005, 0.1]$, the overall noise is low and the performance of the model is expected to be good. However, it should be noted that when $\alpha = [0.005, 0.5]$, SDDPM performs better than all other models. It should also be noted that the performance scores are better in the case of the ultrasound images compared to all the other models. Moreover, Fig. \ref{sar_lines_texture} and \ref{US_denoised} show that even in the case of extreme noise conditions, SDDPM can perform very well. For both datasets, SDDPM can denoise images corrupted with high noise levels while recovering most of the structure present in the original images. 

\subsection{Effect of timestep $T$}

In the case of denoising real images, the exact amount of noise that is present in an image is unknown. It is also almost certain that the noisy images might have noise that does not exactly correspond to the $\alpha_t$ values that we use for training our model. In such a scenario, the model should be able to map the noisy image to one of the noise levels it has been trained with and thereafter generate the denoised image. This makes the designed model robust and can be used for practical denoising applications. 
\begin{figure}[!htb]
    \centering
    \includegraphics[scale = 0.4]{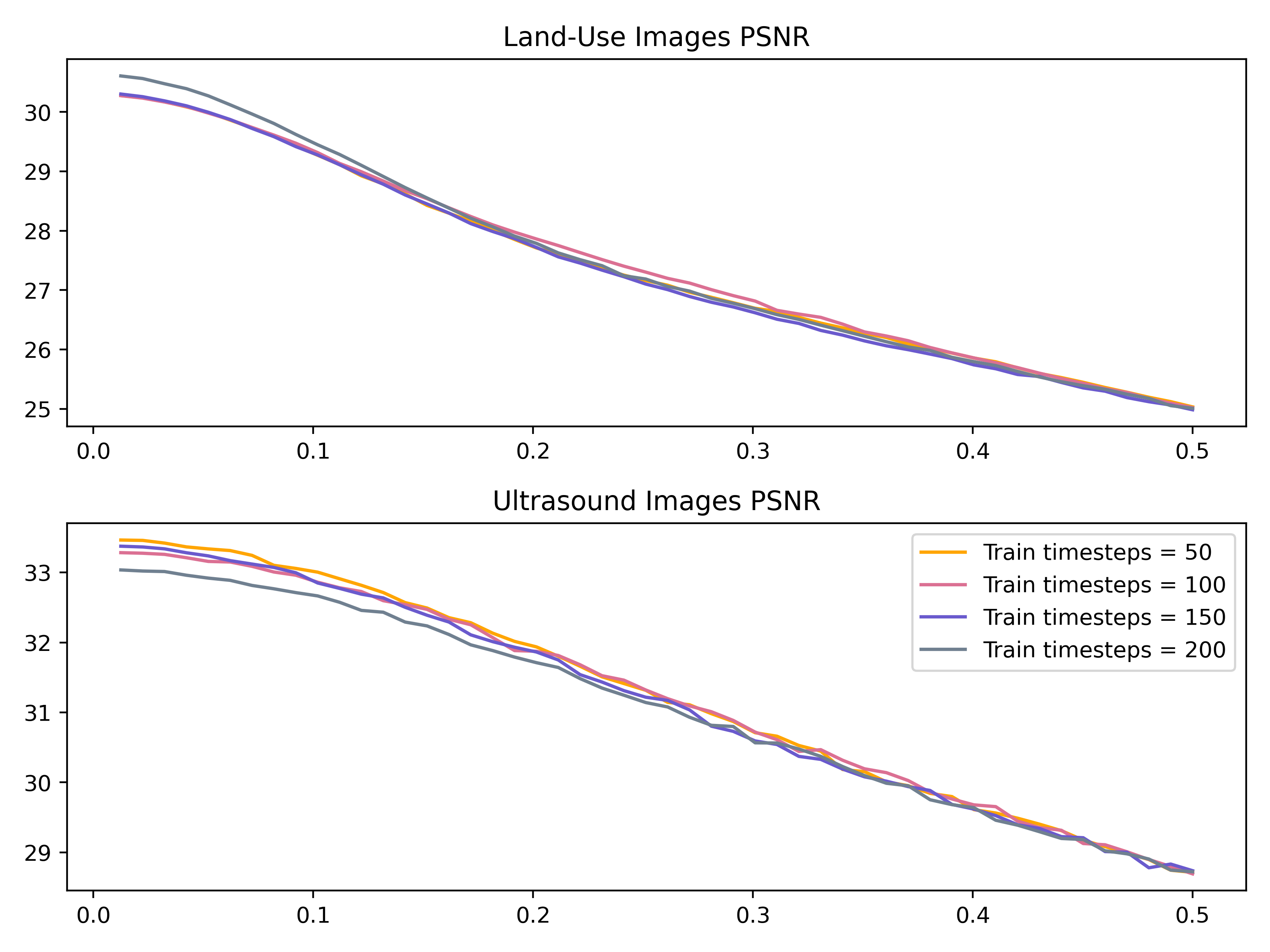}
    \caption{Graphical representations depict the fluctuations in PSNR scores for land-use images in the top row and ultrasound images in the bottom row. These variations occur as a consequence of altering the training timesteps while maintaining the test timesteps at a constant value of 200.}
    \label{psnr_timesteps}
\end{figure}

\begin{figure}[!htb]
    \centering
    \includegraphics[scale = 0.4]{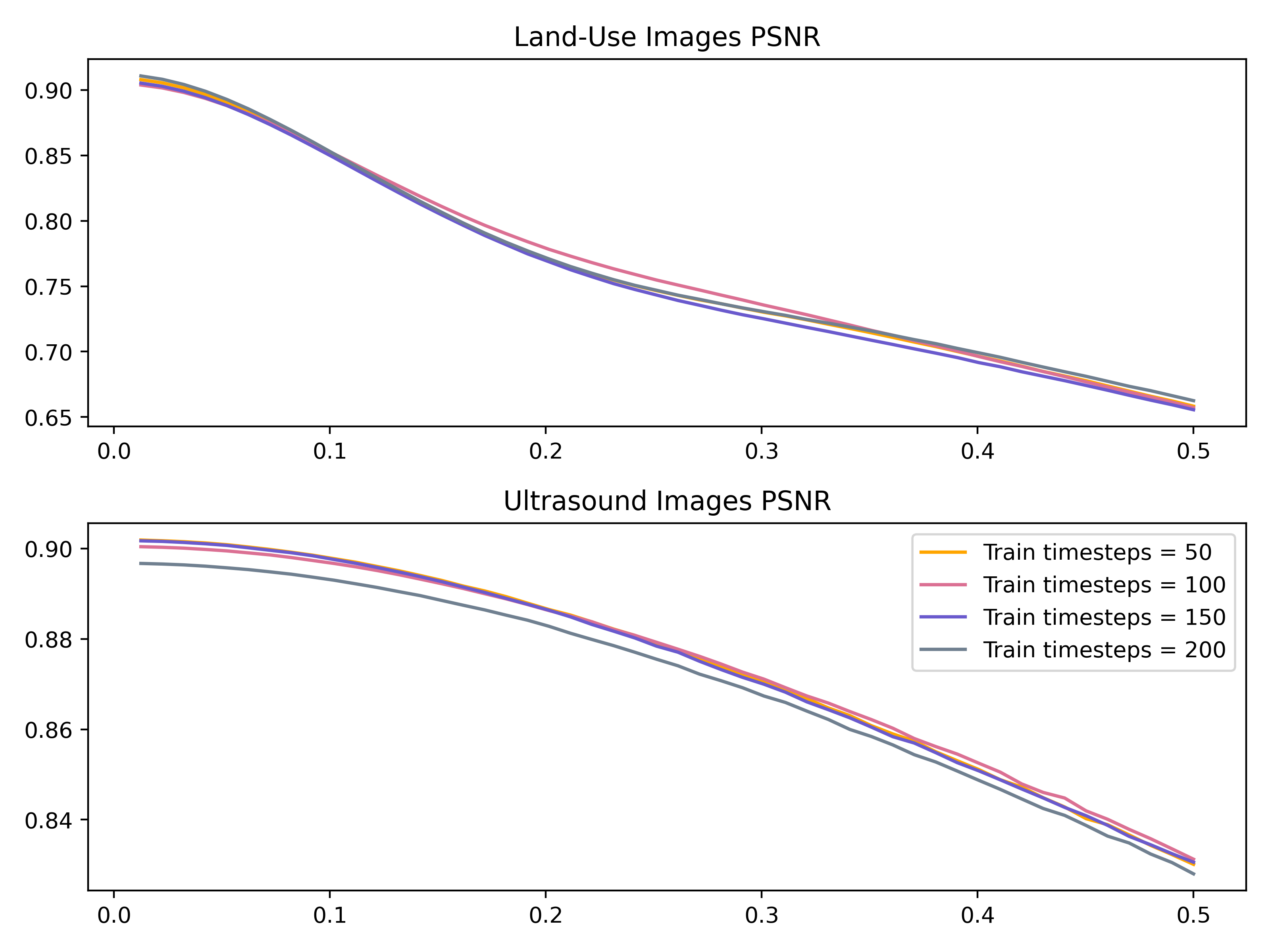}
    \caption{Graphical representations depict the fluctuations in SSIM scores for land-use images in the top row and ultrasound images in the bottom row. These variations occur as a consequence of altering the training timesteps while maintaining the test timesteps at a constant value of 200.}
    \label{ssim_timesteps}
\end{figure}

In order to evaluate the efficiency of SDDPM, we trained the model with different training and testing timesteps. When the training and testing timesteps are exactly the same, the model does not have to interpolate between different values of $\alpha_t$. All the results shown in tables \ref{tab_SAR} and \ref{tab_US} are obtained with the same training and testing timesteps where both were set to $200$ and noise level $\alpha$ was varied between $[0.005, 0.5]$. Fig. \ref{psnr_timesteps} and \ref{ssim_timesteps} show how the mean PSNR and SSIM scores vary when the training timestep is changed keeping the testing timesteps fixed. We fixed the test timesteps to $200$ in all the experiments, trained our model with 50, 100, 150 and 200 timesteps and obtained the PSNR and SSIM scores for both datasets. Each curve shown in Fig. \ref{psnr_timesteps} and \ref{ssim_timesteps} is the mean curve obtained from 4 randomly chosen train and test partitions for a particular pair of training and testing timesteps. It is evident from Fig. \ref{psnr_timesteps} and \ref{ssim_timesteps} that SDDPM is robust to the selection of different training and testing timestep values and can be used for denoising images obtained from the wild. 

\begin{figure}[!htb]
\begin{center}
\includegraphics[width=0.5\textwidth]{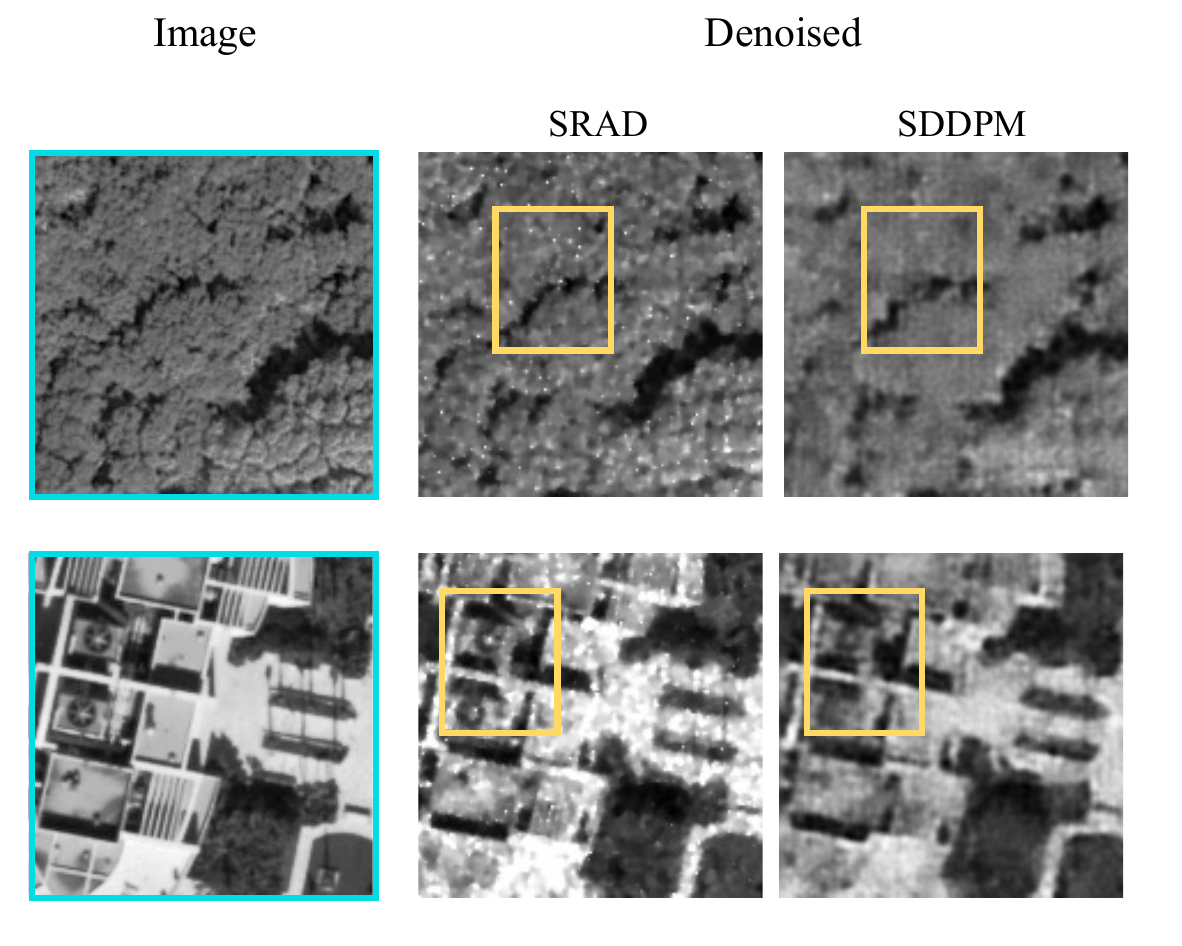}
\end{center}
\caption{Column 1: Patch from the original image. Columns 2 and 3 show the denoised versions obtained by SRAD (column 2) and SDDPM (column 3) for $\alpha_t = 0.45$. }
\label{sradvssddpm}
\end{figure}
\subsection{Performance of SRAD \cite{yu2002speckle} and SDDPM for high noise ($\alpha_t = 0.45$)}

SRAD \cite{yu2002speckle} has been specifically proposed for reducing speckle. 
From Tables \ref{tab_SAR} and \ref{tab_US}, we see that the SSIM scores of SDDPM and SRAD are comparable when noise is low. In scenarios where the noise is high, SDDPM starts performing better than SRAD. This is primarily due to the fact that the proposed model is data-driven and can learn the underlying features which in turn help in recovering the high frequency features. It should also be noted that the PSNR scores of the proposed model are consistently higher than those of SRAD for all noise levels. 

In order to further analyze the denoised images generated by SRAD and SDDPM, we take a closer look at the first two images shown in Fig. \ref{comparison_imgs} for $\alpha_t = 0.45$ in Fig. \ref{sradvssddpm}. The denoised images generated by SRAD and SDDPM for $\alpha_t = 0.45$ have some structural differences. Fig. \ref{sradvssddpm} shows a patch from each of these two denoised images corresponding to $\alpha_t = 0.45$. If we take a closer look inside the yellow boxes in Fig. \ref{sradvssddpm}, we will see that SRAD generates some white speckle artifacts in both images which is not present in the denoised images generated by SDDPM. This \textit{reverse-speckle} phenomenon has been well studied \cite{segall1997morphological} in the context of anisotropic diffusion and is seen in images denoised by SRAD for extreme noise conditions. In contrast, since SDDPM does not have any anisotropic component in the optimizing function, these artifacts are not seen. 

\subsection{Training time, denoising time and parameters}

The computational costs of training and denoising methods play a pivotal role in real-world applicability. The time required to train a model on a dataset and subsequently test its performance can significantly impact its feasibility for deployment in various applications. In this section, we delve into an analysis of the training and test times of DnCNN, SCU, SwinIR and SDDPM.

\begin{figure}[!htb]
\begin{center}
\includegraphics[width=0.5\textwidth]{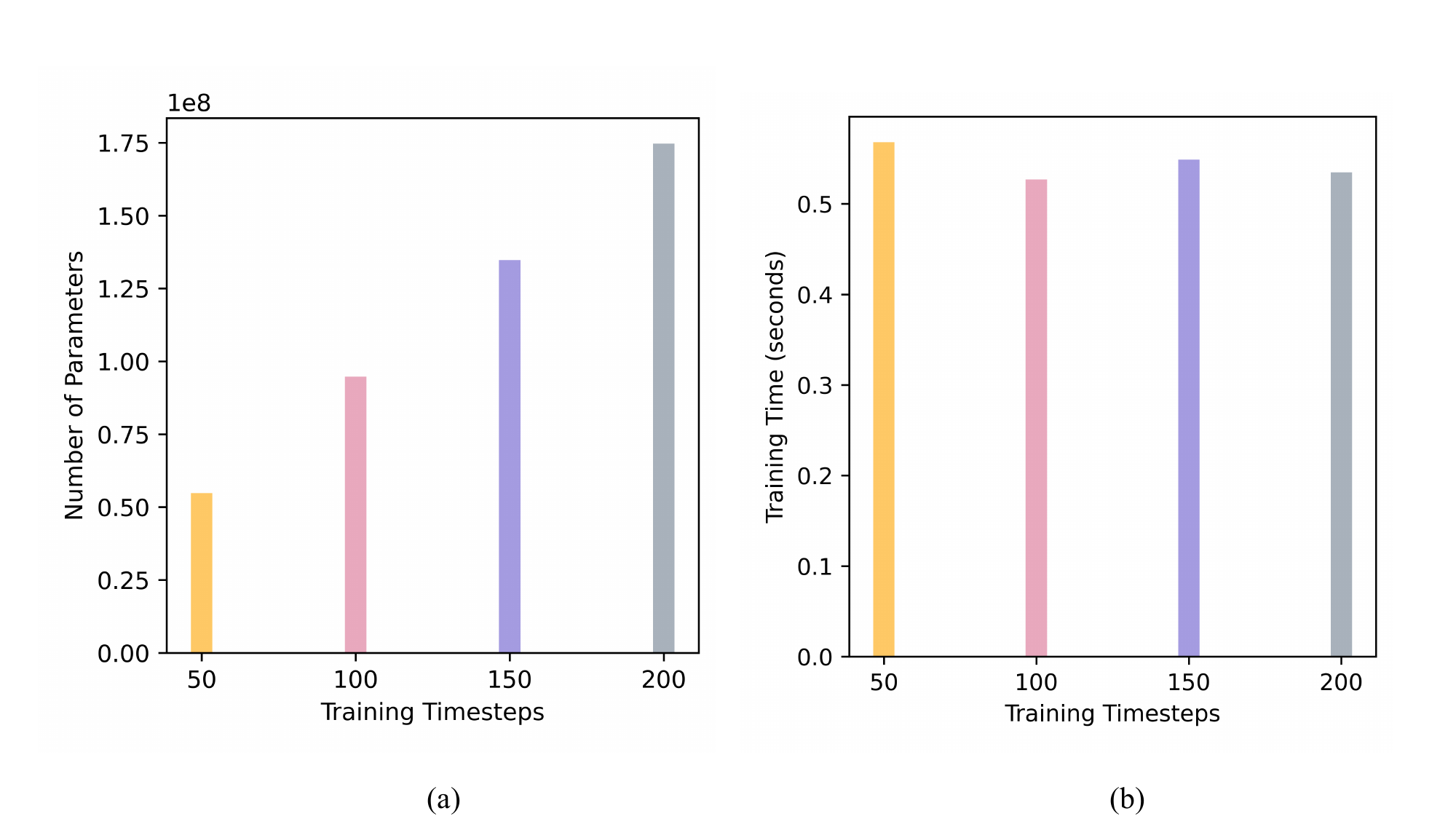}
\end{center}
\caption{(a) This figure shows how the number of parameters of the proposed model SDDPM increases with the increase in training timesteps. (b)  This figure shows how the training time (in seconds) of SDDPM varies with the increase in training timesteps for a batch of images. It can be seen that even though the number of trainable parameters increases with the increase in training timesteps, the training time of a batch of images remains unaffected.}
\label{sddpm_params_traintime}
\end{figure}

\begin{figure}[!htb]
\begin{center}
\includegraphics[width=0.5\textwidth]{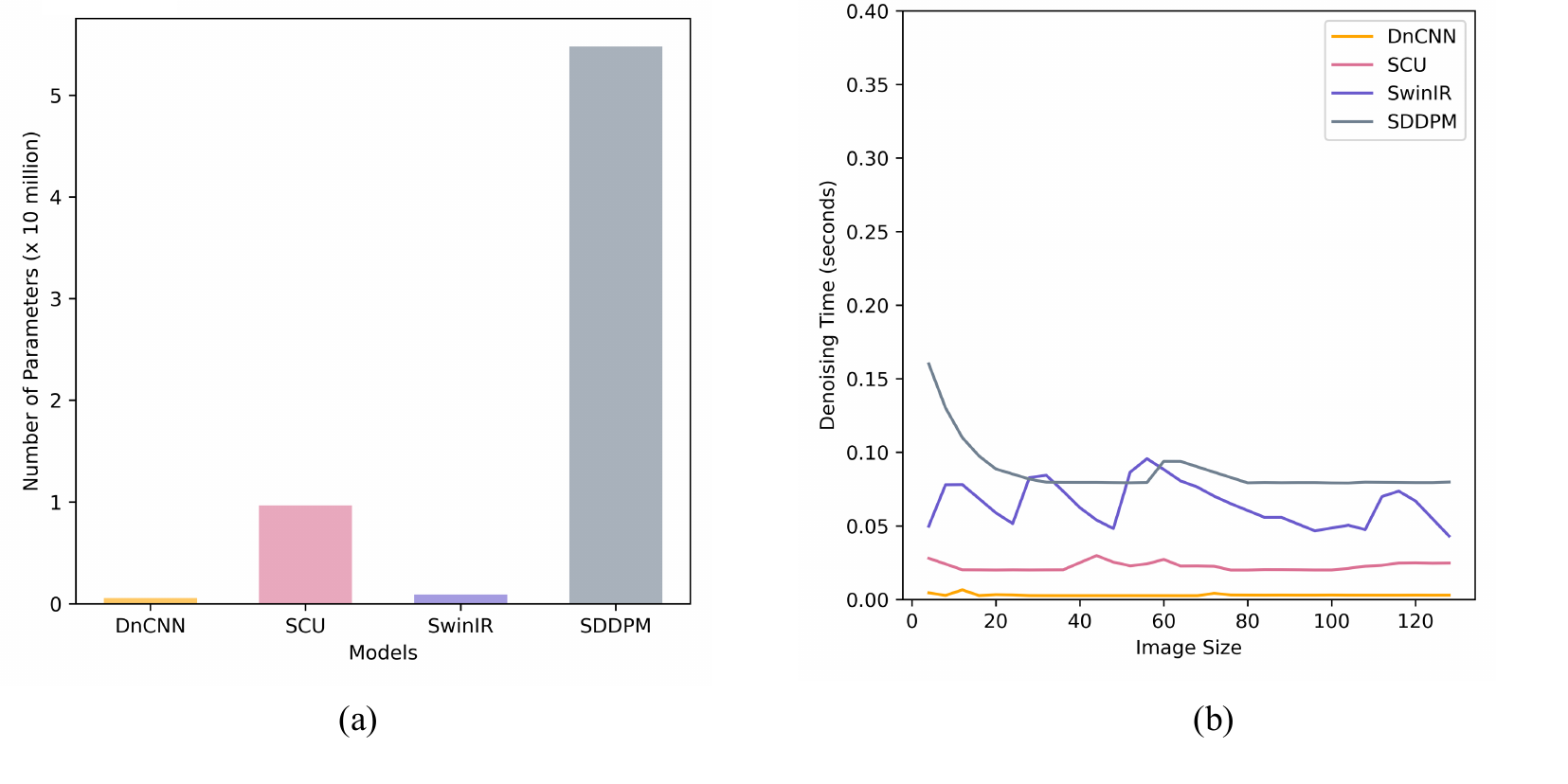}
\end{center}
\caption{(a) This figure shows the number of trainable parameters of DnCNN, SCU, SwinIR and SDDPM. (b)  This figure shows the denoising times (in seconds) of DnCNN, SCU, SwinIR and SDDPM.}
\label{models_params_traintime}
\end{figure}

Fig. \ref{sddpm_params_traintime} shows how the number of trainable parameters of SDDPM increase with the increase in training timesteps and the corresponding training time in seconds. The number of parameters of SDDPM increases monotonically with the increase in training timesteps. This is because the input to the model is $(x_t, \tau)$ where $x_t$ is the noisy image corrupted with noise level $\alpha_t$ and $\tau$ is the noise schedule. SDDPM does not require the explicit knowledge of $t$. Instead, it learns to estimate $t$ from $x_t$ and $\tau$ and then proceeds to denoise the corrupted image accordingly. It should be noted that the increase in model parameters does not affect the training time of SDDPM. This is because $\tau$ is integrated in the initial layers of the model and does not affect the overall training cost significantly. Fig. \ref{models_params_traintime} shows the numbers of parameters and the corresponding denoising (inference) times for SDDPM, DnCNN, SCU and SwinIR. Even though the number of parameters of SDDPM is 10x that of SCU and about 100x that of DnCNN and SwinIR, it should be noted that the denoising time required for one image is not significantly higher than any of the other models for image sizes ranging from 8x8 to 128x128. \\

\subsection{Computational complexity of SRAD and SDDPM}

\begin{figure}[!htb]
\begin{center}
\includegraphics[width=0.4\textwidth]{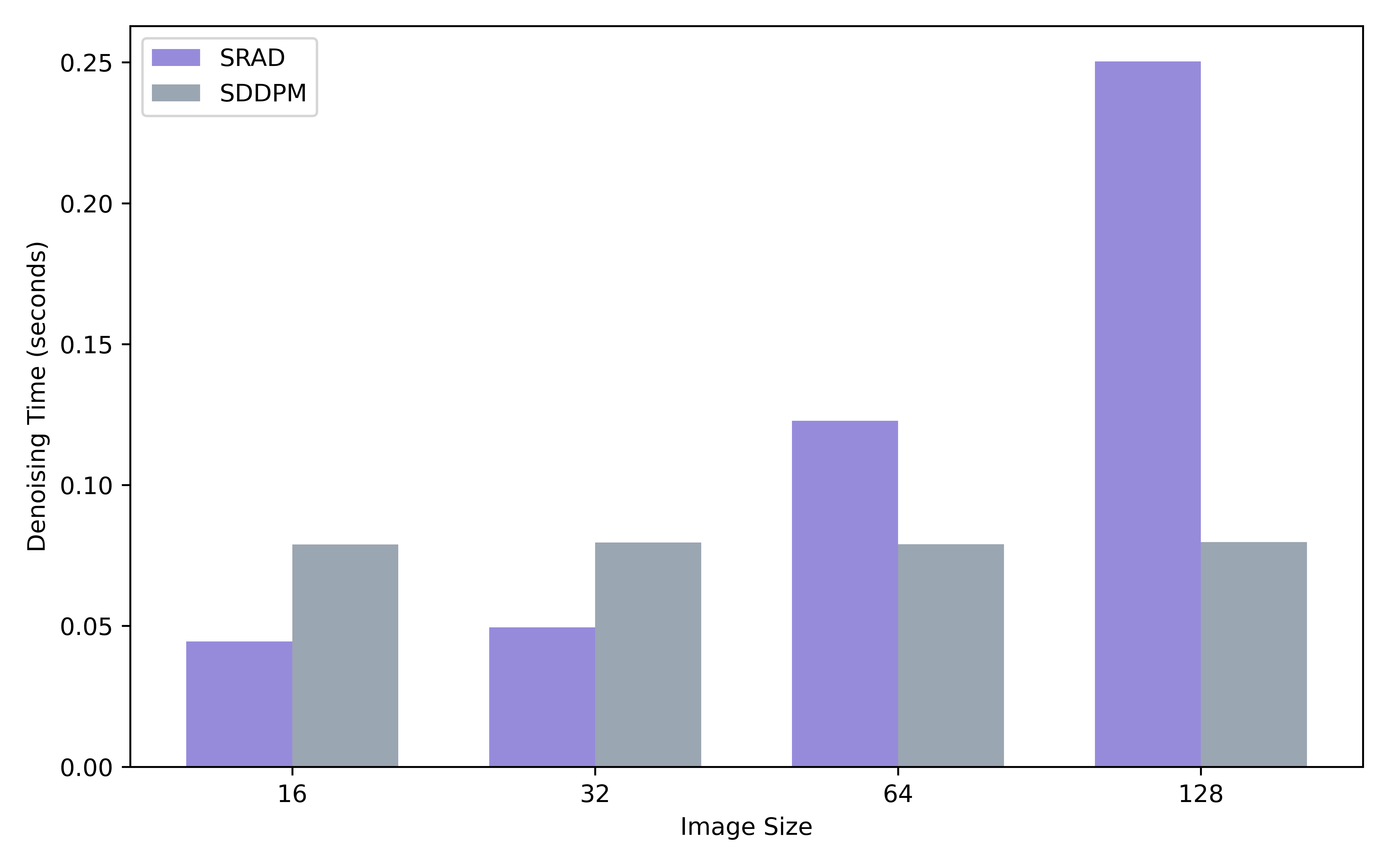}
\end{center}
\caption{This figure shows how the denoising time of SDDPM compares with that of SRAD. SRAD can only denoise one image at a time whereas the time shown for SDDPM corresponds to the denoising time for a batch of 32 images.}
\label{SRADvsSDDPMtime}
\end{figure}

Tables \ref{tab_SAR} and \ref{tab_US} show that SRAD and SDDPM perform consistently even for high noise levels. This is in accordance with our expectations as both the algorithms are particularly designed for reducing speckle. The primary difference between SRAD and SDDPM is that SRAD is iterative, whereas SDDPM is a neural network model and data-driven. As a direct consequence, we can expect the denoising time for SRAD to be higher than SDDPM. Fig. \ref{SRADvsSDDPMtime} shows the comparison between the denoising time in seconds for SRAD and SDDPM. For SRAD, this is the time required for denoising one image whereas for SDDPM, the denoising time shown is for a batch of 32 images. Theoretically, the computational complexity of SRAD can be derived to be $\mathcal{O}(IN^2k^2)$ where $I$ is the total number of iterations, the image is $N \times N$ in size and $k \times k$ is the neighborhood size. For the number of iterations and the same neighborhood size, Fig. \ref{SRADvsSDDPMtime} shows that the denoising time for a single image increases quadratically with the increase in image size. The computational complexity of neural network models depends on a wide range of factors which include the number of layers, the number of convolutional kernels, the kernel sizes and the number of feature maps among various other factors. Though the theoretical computational complexity of SDDPM is difficult to obtain, it can be seen from Fig. \ref{SRADvsSDDPMtime} that the overall denoising time for the trained SDDPM model is constant for a batch of images, i.e, the computational complexity of the denoising (inference) step is $\mathcal{O}(1)$. From Fig. \ref{SRADvsSDDPMtime}, it seems that the denoising time for SDDPM is more than that of SRAD for smaller images ($16 \times 16$) and ($32 \times 32$). However, it should be noted that the denoising time for SDDPM corresponds to the time required for processing a batch of images (in our case 32) and the average denoising time required by SDDPM for a single image is always lower than that required by SRAD for all image sizes. 

\section{Conclusion}

In conclusion, this paper presents a novel diffusion model for removing speckle. The work represents the first development of a forward diffusion process and reverse diffusion process that assumes a multiplicative noise model. In addition, the SDDPM establishes a training objective and process for the speckle removal process. The proposed model SDDPM has been compared to classical image denoising algorithms, including SRAD, BM3D and NLMeans, as well as to three recent convolutional neural network-based models. Extensive experiments on different datasets show that SDDPM outperforms classical and learning-based solutions for almost all noise levels. It has also been shown that SDDPM is capable of recovering images even in case of extreme noise conditions and is robust to the choice of training and test timestep. The diversity among the land-use images used in our model evaluation, spanning 21 distinct categories, underscores the versatility of our proposed model. Furthermore, SDDPM demonstrates notable efficiency in reconstructing structures within noisy ultrasound images. SDDPM showcases its efficacy in denoising a wide range of image types, even when noise levels are exceptionally high. Going forward, we will explore the integration of generative modeling into the denoising process. This avenue of research holds promise, particularly in scenarios involving substantial noise corruption, where traditional denoising algorithms face formidable challenges in accurately reconstructing the original image.

\section*{Acknowledgment}

We extend our heartfelt gratitude to Dr. Jie Wang (Nanjing University of Posts and Telecommunications) and Dr. Matthew Korban (University of Virginia) for their invaluable insights and thoughtful feedback. We also wish to express our gratitude to Tanjin Taher Toma (University of Virginia) for her unwavering support and engaging discussions.



%





\ifCLASSOPTIONcaptionsoff
  \newpage
\fi





\bibliographystyle{IEEEtran}
\bibliography{MTT_reveyrand}
%


\begin{IEEEbiography}[{\includegraphics[width=1in,height=1.25in,clip,keepaspectratio]{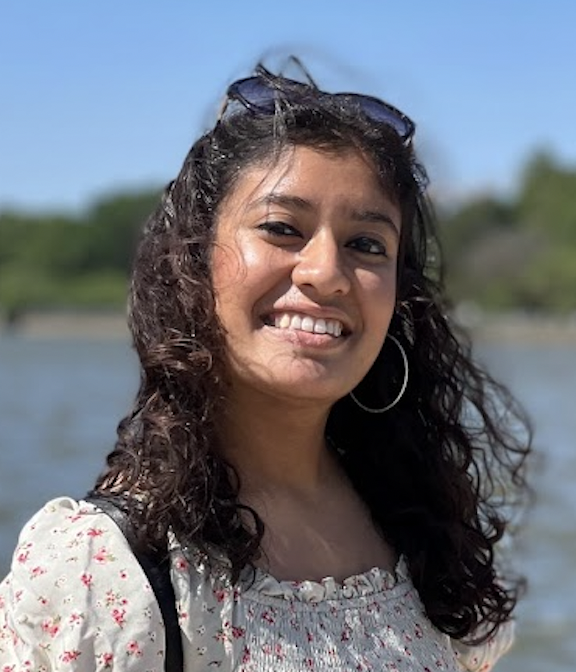}}]{Soumee Guha}
(Student Member, IEEE) received Bachelor of Engineering
in electrical engineering from Jadavpur University, India
in 2018, and MTech in computer science from Indian Statistical Institute in 2020. She is currently pursuing the Ph.D. degree with the Virginia Image and Video Analysis Laboratory (VIVA), University of Virginia. Her current research interests include image processing, biological and biomedical image analysis, image generation, image denoising and deep learning.
\end{IEEEbiography}
\begin{IEEEbiography}[{\includegraphics[width=1in,height=1.25in,clip,keepaspectratio]{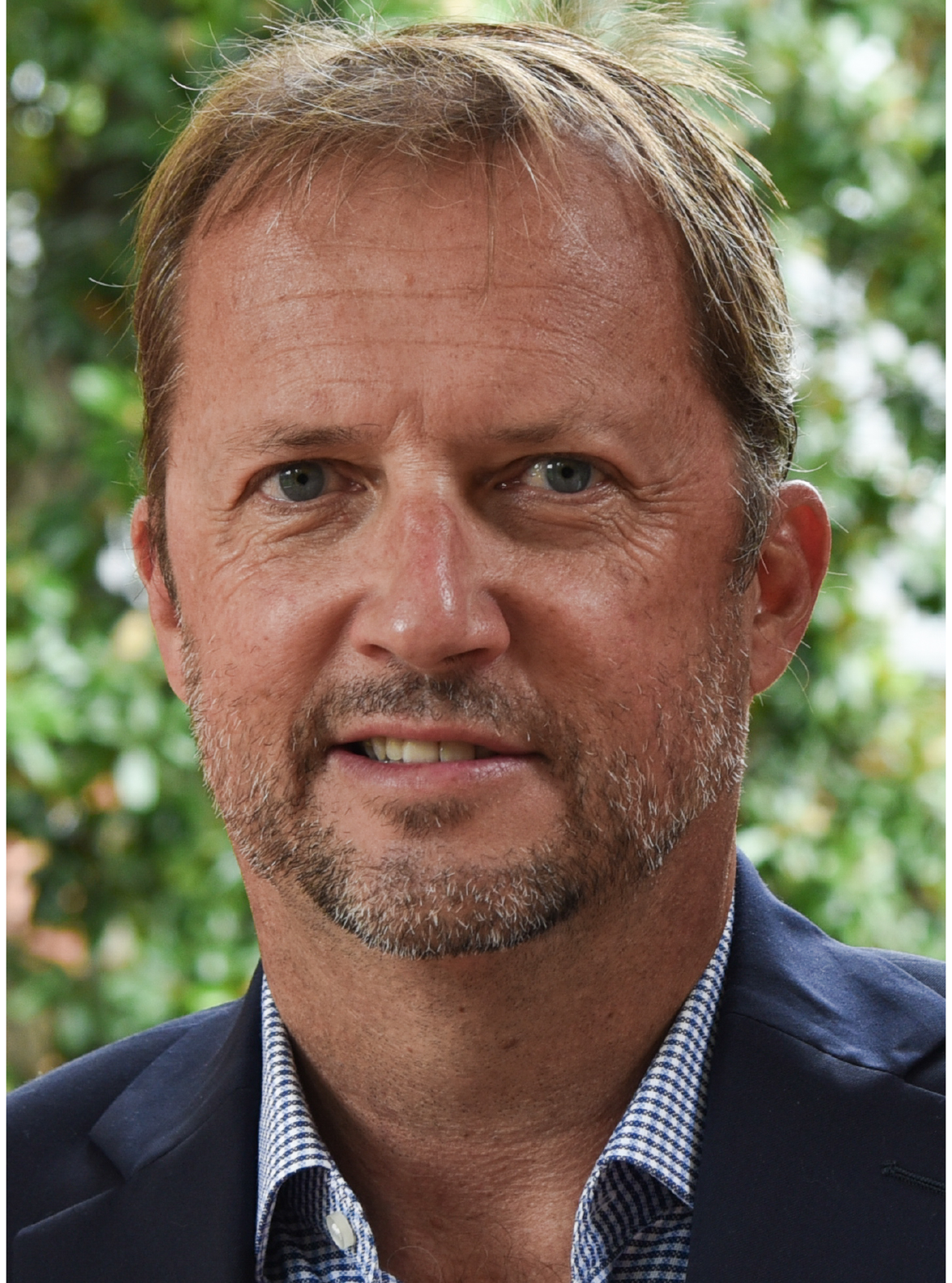}}]{Scott T. Acton}
(Fellow, IEEE) is the Lawrence R. Quarles Professor and Chair of Electrical \& Computer Engineering at the University of Virginia. He is also appointed in Biomedical Engineering. For the previous three years, he was Program Director in the Computer and Information Sciences and Engineering directorate of the National Science Foundation. He received the M.S. and Ph.D. degrees at the University of Texas at Austin, and he received his B.S. degree at Virginia Tech. Professor Acton is a Fellow of the IEEE “for contributions to biomedical image analysis.” Professor Acton’s laboratory at UVA is called VIVA - Virginia Image and Video Analysis. They specialize in biological/biomedical image analysis problems. Professor Acton has over 300 publications in the image analysis area including the books \textit{Biomedical Image Analysis: Tracking and Biomedical Image Analysis: Segmentation}. He was the 2018 Co-Chair of the IEEE International Symposium on Biomedical Imaging. Professor Acton was Editor-in-Chief of the IEEE Transactions on Image Processing (2014-2018).
\end{IEEEbiography}

\vfill


\end{document}